**Huge anisotropic magneto-thermal switching in high-purity polycrystalline compensated metals**

Poonam Rani[1], Yuto Watanabe[1], Takuma Shiga[2], Yuya Sakuraba[3], Hikaru Takeda[4], Minoru Yamashita[4], Ken-ichi Uchida[3,5], Aichi Yamashita[1], Yoshikazu Mizuguchi[1]*

[1] Department of Physics, Tokyo Metropolitan University, Hachioji 192-0397, Japan.

[2] National Institute of Advanced Industrial Science and Technology (AIST), Tsukuba, Ibaraki 305-8568, Japan.

[3] National Institute for Materials Science (NIMS), Tsukuba, Ibaraki 305-0047, Japan.

[4] Institute for Solid State Physics, The University of Tokyo, Kashiwa 277-8581, Japan.

[5] Department of Advanced Materials Science, Graduate School of Frontier Sciences, The University of Tokyo, Kashiwa 277-8581, Japan.

* Corresponding author: Yoshikazu Mizuguchi (mizugu@tmu.ac.jp)

**Abstract**

Magneto-thermal transport is a promising physical property for thermal management applications. Magneto-thermal switching enables active control of heat flows, and a high switching ratio is desirable for improving performance. Here, we report on the observation of a huge magneto-thermal switching (MTS) effect in high-purity (5N) Pb polycrystalline wires, where magnetic fields perpendicular to the heat current direction are applied at low temperatures. At $T = 3$ K and $B = 0.1$ T, the measured thermal conductivity ($\kappa$) of the Pb wire is about 2500 W m$^{-1}$ K$^{-1}$ but is reduced to ~150 and ~5 W m$^{-1}$ K$^{-1}$ at $B = 1$ and 9 T, respectively. This strong suppression is attributed to magnetoresistance in compensated metals. Although the huge magnetoresistance has been studied in single crystals with field along the selected orbitals, our results demonstrate that a huge MTS can similarly be realized even in flexible polycrystalline wires. This finding highlights the practical potential of magneto-thermal control in low-temperature thermal management, including applications in space environments where temperatures are around 3 K.

**Highlights**

- Huge magneto-thermal switching is observed in high-purity polycrystalline Pb wires.

- Polycrystalline wires of compensated metals show high anisotropic magnetoresistance.

- Huge magneto-thermal switching in flexible metal wires is valuable for applications.

**1. Introduction**

Thermal management is a crucial for creating new application and improving its performance [1–4]. Control of heat flow enhances efficiency of energy storage and device performance. Among various thermal management strategies, magneto-thermal switching (MTS) is particularly attractive because it enables achieve heat flow control without any mechanical motions [5,6]. Recently, we demonstrated that a large MTS ratio, that is defined as the difference in thermal conductivity ($\kappa$) with/without an applied magnetic field ($B$), can be achieved in pure-element superconductors including Nb and Pb with high purity [7–9]. In addition, we have shown that phase-separated alloys, including Sn-Pb and In-Sn solders, exhibit nonvolatile MTS behavior [10,11]. Although MTS effect in superconductors emerges at low temperatures (lower than superconducting transition temperatures), development of low-temperature MTS materials is useful for applications such as quantum computing, adiabatic demagnetization refrigerators, and space technologies, where the average temperature is about 3 K [12,13]. Furthermore, MTS effect for pure metals in normal-conducting states have been extensively studied using single crystals of compensated metals because of strong reduction in thermal conductivity caused by magnetoresistance; as an example for tungsten (W) single crystal with $B$ along the [001] direction, it has been reported that the electronic contribution of $\kappa$ ($\kappa_{el}$) can be almost completely suppressed [14–17]. Although the reduction in $\kappa_{el}$ for compensated metals typically occurs at low temperatures, the MTS ratio can be remarkably high on the order of 1000000% in high-purity W single crystals [18]. There has been a study on polycrystalline samples of high-purity aluminum (Al), where the observed MTS ratio is lower than 600% [19], but the studies on MTS in high-purity polycrystalline metal samples have been limited. Here, we report on the observation of huge magnetoresistance and MTS effect in high-purity (5N) Pb polycrystalline wires; Pb is a compensated metal, because of an even number of electrons per unit cell. The observation of a huge MTS in randomly oriented polycrystalline wires is the major finding of this study. Although the observed MTS ratio was enhanced at lower temperatures, the inherent shape flexibility of polycrystalline metals is beneficial for thermal management in low-temperature applications.

2. **Experimental details**

The investigated Pb polycrystalline wires (5N purity) with a diameter of 0.5 mm (Fig. S1) was purchased from The Nilaco Corporation. All the measurements have been done within two months after the first exposure to the air to avoid oxidation effects on the samples. As shown in Supplemental Materials (Fig. S2), thermal transport properties of the Pb (5N) wires kept in air have been slightly degraded in two years, which may be caused by surface oxidation. High-purity

Al (5N purity) and Zn (5N purity) polycrystalline wires with a diameter of 0.5 mm were also purchased from The Nilaco Corporation for a comparative study.

Measurements of $\kappa$ and electrical resistivity ($\rho$) with four-terminal methods were performed on Physical Property Measurement System (PPMS-Dynacool, Quantum Design) using the thermal transport option (TTO) and the DC resistivity option, respectively. For the $\kappa$ measurements, the terminals were fabricated using Ag paste and Cu wires with a diameter of 0.2 mm, and the field direction was controlled by changing the sample setup. For the $\rho$ measurements, the terminals were fabricated using Ag paste and Au wires with a diameter of 0.025 mm, and a rotator probe was used to control the field direction.

3. Results

Figure 1 summarizes the thermal transport properties of the high-purity Pb (Pb-5N, polycrystalline, sample #1) wires. $\kappa$ was measured under two magnetic field ($B$) orientations relative to the heat current ($J$), $\boldsymbol{B} \perp \boldsymbol{J}$ and $\boldsymbol{B} // \boldsymbol{J}$, as shown in Figs. 1(a) and 1(b). For $\kappa$ measurements with $\boldsymbol{B} \perp \boldsymbol{J}$, the sample was bent into a W-like shaped geometry to increase the terminal distance, which is necessary for accurately measuring high $\kappa$ values. Figure 1(c) shows the temperature dependence of $\kappa$ ($\kappa$-$T$) at magnetic fields of 0, 0.04, and 0.08 T. In the normal conducting states at 0.08 T, $\kappa$ increases with decreasing temperature, consistent with a previous report [8]. Figures 1(d) and 1(e) show the magnetic field dependence of $\kappa$ ($\kappa$-$B$) measured at $T$ = 2 K (with $\boldsymbol{B} \perp \boldsymbol{J}$) and 3 K ($\boldsymbol{B} \perp \boldsymbol{J}$ and $\boldsymbol{B} // \boldsymbol{J}$), respectively. As shown in Fig. 1(e), $\kappa$ is more strongly suppressed by the field of $\boldsymbol{B} \perp \boldsymbol{J}$ than that of $\boldsymbol{B} // \boldsymbol{J}$. A large MTS was observed in all the different Pb-5N wires (Sample #1, #2, and #4; see Figs. 3 and S3), whereas low-purity (3N) wires exhibited no such behavior (Fig. S3). Figure 2 shows the magnetic field dependence of the magneto-thermal switching ratios (MTSRs) defined as MTSR = ($\kappa_{max}$-$\kappa$)/ $\kappa$, where $\kappa_{max}$ denotes the highest $\kappa$ observed at above the critical field. At $T = 2$ K, the MTSR exceeds 80000%: for example, $\kappa$ at $B = 9$ T is >800 times lower than that at $B \sim 0.1$ T as shown in Fig. 2(a). In Fig. 2(b), we see that the MTSR ($\bm{B} \perp \bm{J}$) at $T = 3$ K is lower than that at $T = 2$ K, but the ratio is still high and exceeding 45000%. The large MTS observed under the fields of $\bm{B} \perp \bm{J}$ can be attributed to the magnetoresistance of compensated metals, but the observation of such huge MTS effect in a polycrystalline wire was unexpected. Hereafter, we show some transport properties of various samples to demonstrate that the observed large MTS is originated from the magnetoresistance of compensated metals. As mentioned in the introduction part, previous studies for the magnetoresistance of compensated metals have primarily focused on single crystals with selected field direction. Therefore, we need to find additional evidences for that the magnetoresistance in compensated metals can also emerge in randomly formed polycrystalline wires, and we measured the angle dependence of magnetoresistance.

We measured $\rho$ and $\kappa$ of the same Pb-5N sample (sample #2) with a similar W-like shape with magnetic fields perpendicular to electrical current and heat flow, which is shown in Fig. 3. When switching the terminal configurations from $\rho$ to $\kappa$ measurements, the sample shape was maintained as much as possible. Figures 3(a) and 3(b) show the magnetoresistance ($\rho$-$B$) for sample #2. The obtained $\rho$ is nearly proportional to $B^2$, which is a typical characteristic of compensated metals. At lower fields, we see superconducting transition, and the observed $\rho$ in the normal-conducting states at around the critical field is extremely low, which yields high $\kappa$ in the normal-conducting states at around the critical field. Figure 3(c) displays the $\kappa$-$B$ at $T = 3$ K measured after the $\rho$ measurements. In Fig. 3(d), the comparison of the experimental values of total $\kappa$ and the theoretical values of $\kappa_{el}$ is shown. Here, the theoretical values of $\kappa_{el}$ were estimated through the Wiedemann-Franz law with the measured $\rho$, namely from the relationship of $\kappa\rho = LT$, where $L$ is the Lorenz number. At lower fields, the estimation of $\kappa_{el}$ was less accurate due to the difficulty of the $\rho$ measurements by the small $\rho$, but that at $B > 1$ T was reliable because of the large magnetoresistance. As shown in Fig. 3(d), the difference between the experimental values of total $\kappa$ and the theoretical values of $\kappa_{el}$ converges to about 4-5 W m$^{-1}$ K$^{-1}$, which likely corresponds to the phonon contribution. The order of the phonon thermal conductivity is consistent with previous works on high-purity metals [20,21]. The results suggest that the $\kappa_{el}$ is mostly suppressed by the field of $\boldsymbol{B} \perp \boldsymbol{J}$ at $B = 9$ T.

Next, we investigated the angle dependence of magnetoresistance using a rotator system, where the straight-shaped sample was used (Fig. 4(a)). Figures 4(b), 4(c), and 4(d) show the angle dependence of $\rho$ at $B = 9$, 1, and 0.1 T, respectively. Here, the angles of 0 and 90 degree correspond to the perpendicular-field configuration and parallel-field to the current direction, respectively. Both temperature and magnetic field strength should influence the angle-dependent magnetoresistance. As for temperature, the angle dependence was clearly observed at $T = 3$ and 4 K, but it was largely suppressed at $T = 10$ K under $B = 1$ and 9 T. Regarding magnetic field strength, the angle dependence disappears at $B = 0.1$ T. These observations support the interpretation that the anisotropic MTS effect observed in high-purity Pb originates from the magnetoresistance of compensated metals under different field orientations.

## 4. Discussion

If phonons predominantly contribute to $\kappa$ at the high field (~ 9 T), the temperature dependence of $\kappa$ is expected to show a $T^3$ temperature dependence at low temperatures. Therefore, we measured $\kappa$-$T$ of Pb-5N wire at various $B$ in $\boldsymbol{B} \perp \boldsymbol{J}$ configuration (Fig. S5(a)). Figures S5(b-d) show the $\kappa$-$T^3$ plots with eye guide lines. It is clear that the temperature range, where the $\kappa$ is proportional to $T^3$, becomes wider with increasing $B$. As shown in Fig. S5(d), $\kappa$ at $B = 9$ T shows the $T^3$ temperature dependence below 6 K, which suggests that $\kappa$ at low temperatures with $B = 9$ T is dominated by phonon contributions.

To further confirm that the observed large MTSR in a polycrystalline Pb-5N wires with fields ($\boldsymbol{B} \perp \boldsymbol{J}$) is caused by the magnetoresistance observed in single crystals of compensated metals, $\kappa$-$B$ for a non-compensated metal Al-5N wire and a compensated metal Zn-5N wire were measured (Figs. S6 and S7). As shown in Fig. S6, although Al does not show a clear difference between the data with $\boldsymbol{B} \perp \boldsymbol{J}$ and $\boldsymbol{B} // \boldsymbol{J}$, large MTSRs are observed in Zn with $\boldsymbol{B} \perp \boldsymbol{J}$ (Fig. S7), which is qualitatively similar to the results of compensated metal Pb wires.

Here, we reported the large MTSR with magnetic fields ($\boldsymbol{B} \perp \boldsymbol{J}$) in polycrystalline wires of compensated metals Pb and Zn with high purity (5N grade). The earlier works on magnetoresistance of compensated metals have mainly focused on single crystal samples with magnetic field along the selected (high symmetric) directions. However, the current samples are randomly oriented polycrystalline ones. Therefore, the use of polycrystalline flexible wires of compensated metals is potential for magneto-thermal management devices with complicated structure.

## 5. Summary

We investigated magneto-thermal transport and magnetoresistance of high-purity (5N grade) metal wires of Pb, Al, and Zn. For Pb and Zn wires, a large magneto-thermal switching ratio exceeding 80000% has been observed with $\boldsymbol{B} \perp \boldsymbol{J}$ because of the nature of compensated metals. From analyses of magnetoresistance, we confirmed that the electronic contribution of $\kappa$ is mostly suppressed at $B = 9$ T ($\boldsymbol{B} \perp \boldsymbol{J}$) for Pb. Furthermore, angle dependence of magnetoresistance confirmed that the conditions of the observation of large MTS in Pb-5N wire are low temperatures and large fields, which is consistent with the magnetoresistance of compensated metals. The high-purity polycrystalline wires of compensated metals will be useful for low-temperature magneto-thermal management.

**Acknowledgements**

This work was partly supported by JST-ERATO (No.: JPMJER2201) and TMU research funds for young scientists.

**Data Availability**

All the data presented in this article can be provided by reasonable requests to the corresponding author.

**References**

[1] J. Jia, S. Li, X. Chen, Y. Shigesato, Emerging Solid–State Thermal Switching Materials, *Adv. Funct. Mater.* **34**, 2406667 (2024). https://doi.org/10.1002/adfm.202406667.

[2] N. Li, J. Ren, L. Wang, G. Zhang, P. Hanggi, B. Li, Phononics: Manipulating heat flow with electronic analogs and beyond, *Rev. Mod. Phys.* **84**, 1045 (2012). https://doi.org/10.1103/RevModPhys.84.1045.

[3] G. Wehmeyer, T. Yabuki, C. Monachon, J. Wu, C. Dames, Thermal diodes, regulators, and switches: Physical mechanisms and potential applications, *Appl. Phys. Rev.* **4**, 041304 (2017). https://doi.org/10.1063/1.5001072.

[4] Q. Zheng, M. Hao, R. Miao, J. Schaadt, C. Dames, Advances in thermal conductivity for energy applications: a review, *Prog. Energy* **3**, 012002 (2021). https://doi.org/10.1088/2516-1083/abd082.

[5] J. Kimling, J. Gooth, K. Nielsch, Spin-dependent thermal transport perpendicular to the planes of Co/Cu multilayers, *Phys. Rev. B* **91**, 144405 (2015). https://doi.org/10.1103/PhysRevB.91.144405.

[6] H. Nakayama, B. Xu, S. Iwamoto, K. Yamamoto, R. Iguchi, A. Miura, T. Hirai, Y. Miura, Y. Sakuraba, J. Shiomi, K. Uchida, Above-room-temperature giant thermal conductivity switching in spintronic multilayers, *Appl. Phys. Lett.* **118**, 042409 (2021). https://doi.org/10.1063/5.0032531.

[7] M. Yoshida, M. R. Kasem, A. Yamashita, K. Uchida, Y. Mizuguchi, Magneto-thermal-switching properties of superconducting Nb, *Appl. Phys. Express* **16**, 033002 (2023). https://doi.org/ 10.35848/1882-0786/acc3dd.

[8] M. Yoshida, H. Arima, A. Yamashita, K. Uchida, Y. Mizuguchi, Large magneto-thermal-switching ratio in superconducting Pb wires, *J. Appl. Phys.* **134**, 065102 (2023). https://doi.org/10.1063/5.0159336.

[9] M. Yoshida, H. Arima, Y. Watanabe, A. Yamashita, Y. Mizuguchi, "Magneto-thermal-switching in type-I and type-II superconductors, *Physica C* **623**, 1354536 (2024). https://doi.org/10.1016/j.physc.2024.1354536.

[10] H. Arima, Md. R. Kasem, H. Sepehri-Amin, F. Ando, K. Uchida, Y. Kinoshita, M. Tokunaga, Y. Mizuguchi, Observation of nonvolatile magneto-thermal switching in superconductors, *Commun. Mater.* **5**, 34 (2024). DOI: 10.1038/s43246-024-00465-9.

[11] P. Rani, T. Murakami, Y. Watanabe, H. Sepehri-Amin, H. Arima, A. Yamashita, Y. Mizuguchi, Nonvolatile magneto-thermal switching driven by vortex trapping in commercial

In-Sn solder, *Appl. Phys. Express* **18**, 033001 (2025). https://doi.org/10.35848/1882-0786/adb6ee.

[12] H. L. Huang, D. Wu, D. Fan, X. Zhu, Superconducting quantum computing: a review. *Sci. China Inf. Sci.* **63**, 180501 (2020). https://doi.org/10.1007/s11432-020-2881-9.

[13] M. J. DiPirro, P. J. Shirron, Heat Switches for ADRs, *Cryogenics* **62**, 172 (2014). http://doi.org/10.1016/j.cryogenics.2014.03.017.

[14] J. M. L. Engels, F. W. Gorter, A. R. Miedema, Magnetoresistance of gallium - a practical heat switch at liquid helium temperatures, *Cryogenics* **12**, 141 (1972). https://doi.org/10.1016/0011-2275(72)90016-1.

[15] E. R. Canavan, M. J. Dipino, M. Jackson, J. Panek, P. J. Shirron, J. G. Tuttle, *AIP Conf. Proc.* **613**, 1183 (2002). https://doi.org/10.1063/1.1472144.

[16] D. K. Wagner, Lattice Thermal Conductivity and High-Field Electrical and Thermal Magnetoconductivities of Tungsten, *Phys. Rev. B* **5**, 336 (1972). https://doi.org/10.1103/PhysRevB.5.336.

[17] J. M. Duval, B. Cain, P. T. Timbie, Design of a Miniature Adiabatic Demagnetization Refrigerator, *ADVANCES IN CRYOGENIC ENGEINEERING: Transactions of the Cryogenic Engineering Conference - CEC, 2002, Anchorage, United States.* pp.1729 (2002). https://doi.org/10.1063/1.1774872.

[18] J. Bartlett, G. Hardy, I. Hepburn, R. Ray, S. Weatherstone, Thermal characterisation of a tungsten magnetoresistive heat switch, *Cryogenics* **50**, 647 (2010). https://doi.org/10.1016/j.cryogenics.2010.02.027.

[19] F. R. Fickett, Magnetoresistance of Very Pure Polycrystalline Aluminum, *Phys. Rev. B* **3**, 1941 (1971). https://doi.org/10.1103/PhysRevB.3.1941.

[20] R. Fletcher, M. R. Stinson, The High-Field Thermal and Electrical Magnetoconductivities of

Pb, *J. Low Temp. Phys.* **22**, 787 (1977). https://doi.org/10.1007/BF00655708.

[21] M. Yao, M. Zebarjadi, C. P. Opeil, Experimental determination of phonon thermal conductivity and Lorenz ratio of single crystal metals: Al, Cu, and Zn, *J. Appl. Phys.* **122**, 135111 (2017). https://doi.org/10.1063/1.4997034.

**Figures**

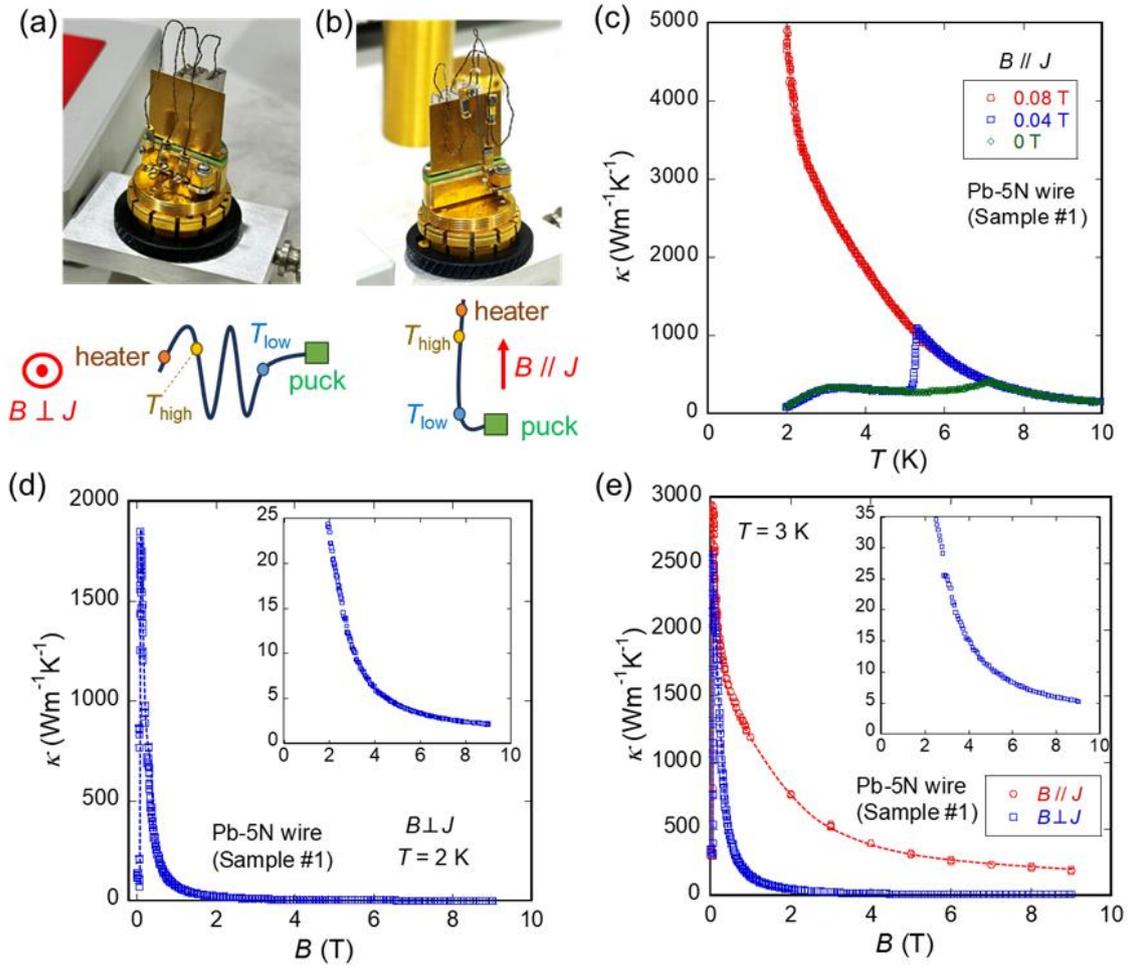

**Fig. 1. Thermal transport measurements of Pb-5N wires.** (a,b) Sample photos and schematic images of sample setup regarding the field direction (***B* ⊥ *J*** and ***B* // *J***). (c) *κ-T* of the Pb-5N wire with ***B* // *J*** under *B* = 0, 0.04, and 0.08 T. (d,e) *κ-B* of the Pb-5N wire at *T* = 2 and 3 K.

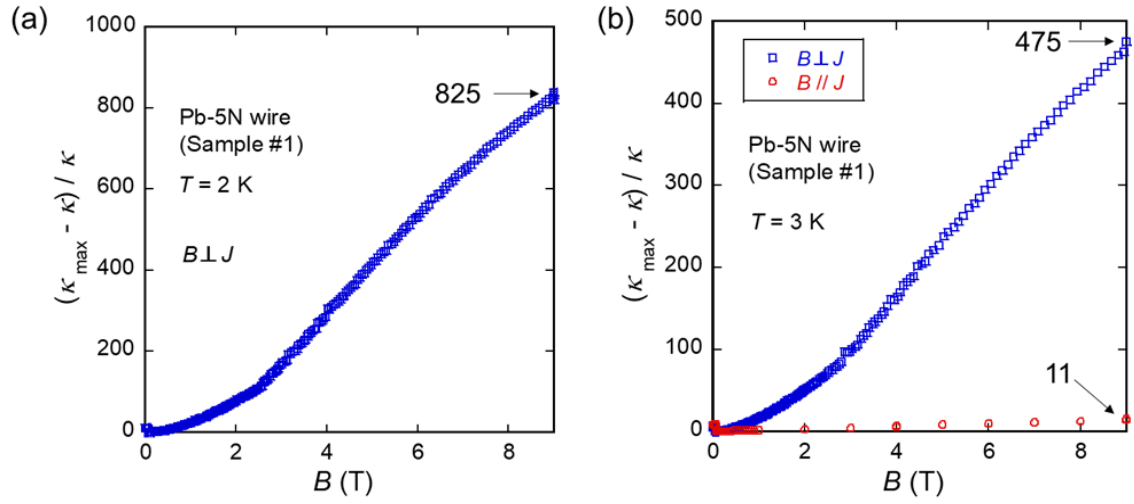

**Fig. 2. Magneto-thermal switching ratio of Pb-5N wires.** (a,b) Field dependence of magneto-thermal switching ratio (MTSR) defined as MTSR = $(\kappa_{max}-\kappa)/\kappa$, where $\kappa_{max}$ denotes the highest $\kappa$ observed at above $H_c$, at $T$ = 2 and 3 K.

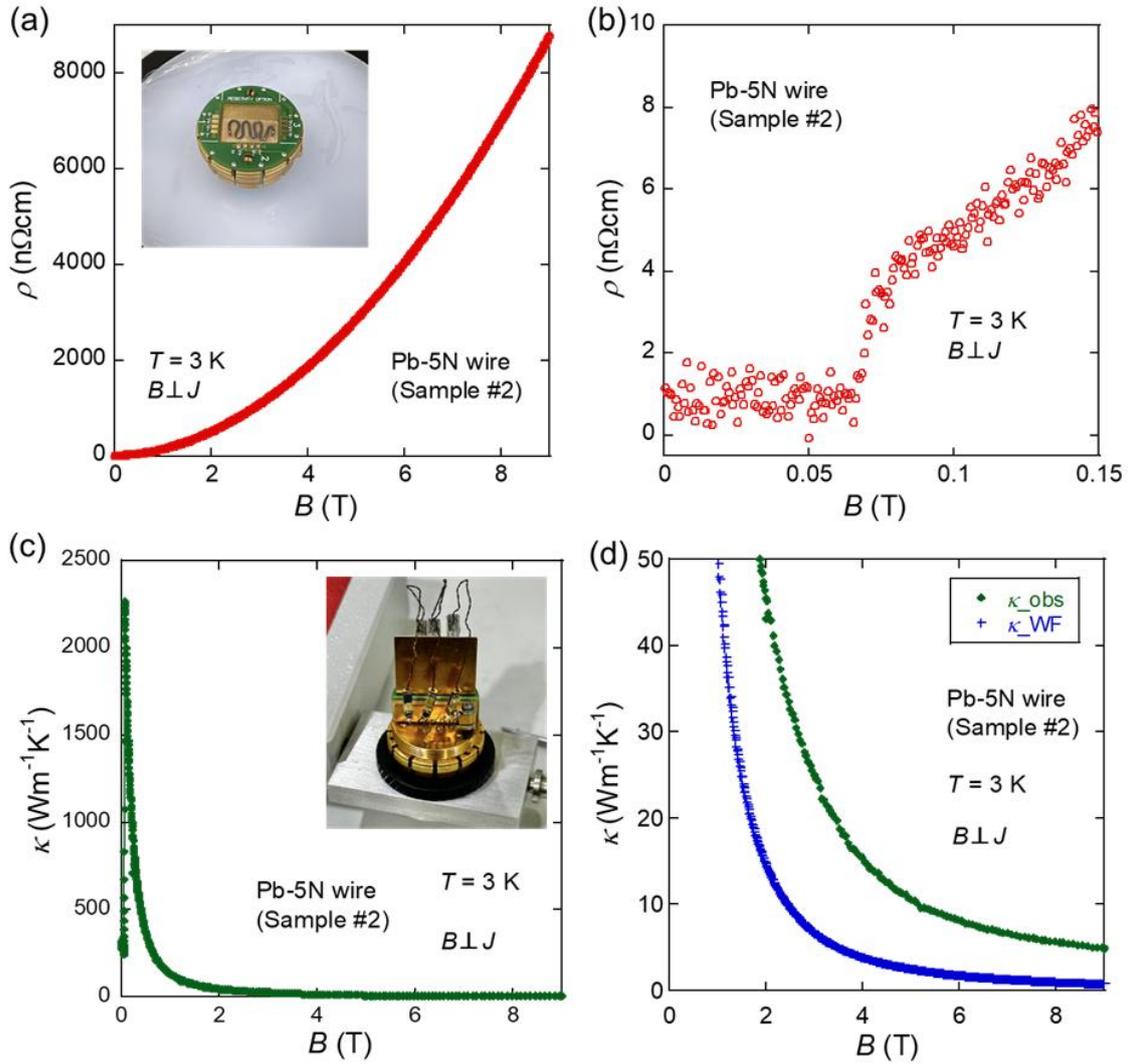

**Fig. 3. Electrical and thermal transport of Pb-5N wire (sample #2).** (a) $\rho$-$B$ of Pb-5N wire and the sample photo (inset) for sample #2. Here, $J$ indicates electrical current direction as well. (b) Low-field data of $\rho$-$B$. (c) Field dependence of $\kappa$ for sample #2 and the sample photo (inset). (d) Comparison of the experimental values of total $\kappa$ ($\kappa$_obs) and the theoretical values of $\kappa_{el}$ estimated by the Wiedemann-Franz law ($\kappa$_WF).

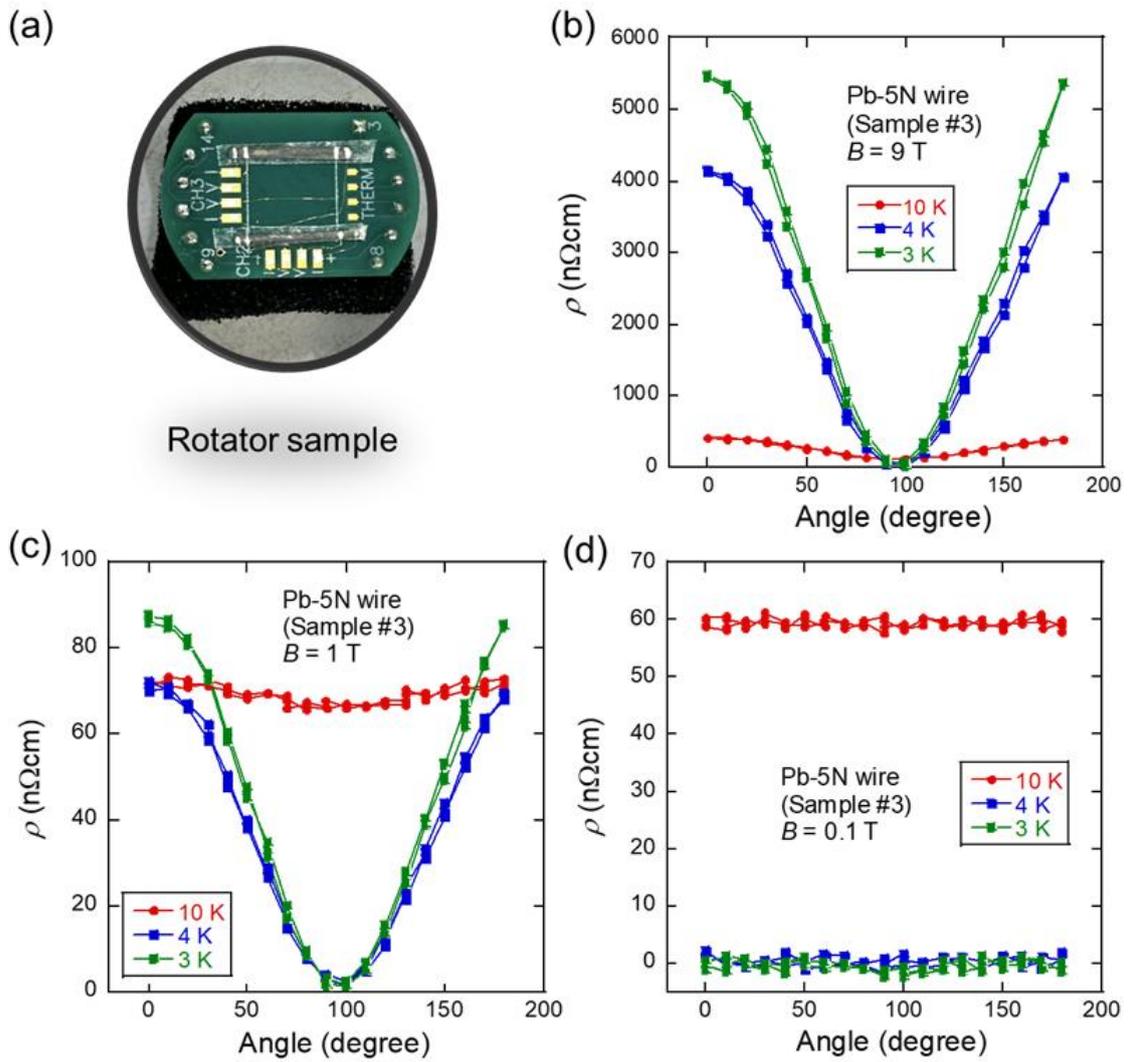

**Fig. 4. Field angle dependence of electrical resistivity of Pb-5N wire (sample #3).** (a) Photo of rotator samples. (b–d) Angle dependence of $\rho$ at (b) $B = 9$ T, (c) $B = 1$ T, and (d) $B = 0.1$ T.

# Supplementary materials

**Huge anisotropic magneto-thermal switching in high-purity polycrystalline compensated metals**

Poonam Rani[1], Yuto Watanabe[1], Takuma Shiga[2], Yuya Sakuraba[3], Hikaru Takeda[4], Minoru Yamashita[4], Ken-ichi Uchida[3,5], Aichi Yamashita[1], Yoshikazu Mizuguchi[1]*

[1] Department of Physics, Tokyo Metropolitan University, Hachioji 192-0397, Japan.

[2] National Institute of Advanced Industrial Science and Technology (AIST), Tsukuba, Ibaraki 305-8568, Japan.

[3] National Institute for Materials Science (NIMS), Tsukuba, Ibaraki 305-0047, Japan.

[4] Institute for Solid State Physics, The University of Tokyo, Kashiwa 277-8581, Japan.

[5] Department of Advanced Materials Science, Graduate School of Frontier Sciences, The University of Tokyo, Kashiwa 277-8581, Japan.

* Corresponding author: Yoshikazu Mizuguchi (mizugu@tmu.ac.jp)

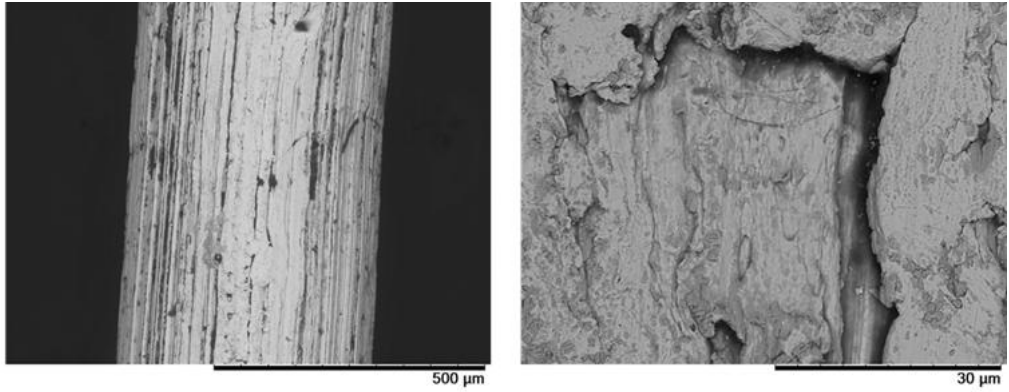

Fig. S1. Scanning electron microscope (SEM) images of the Pb-5N polycrystalline wires. The SEM images were taken on TM3030 (Hitachi High-Tech).

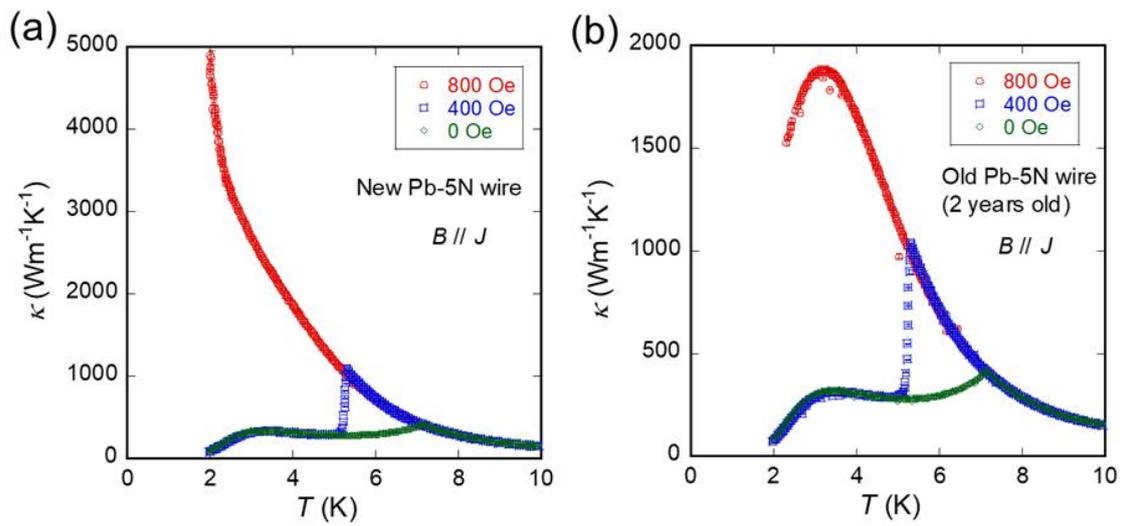

Fig. S2. Comparison of temperature dependences of thermal conductivity ($\kappa$-$T$) for a newly opened Pb-5N wire and an old Pb-5N wire opened two years ago.

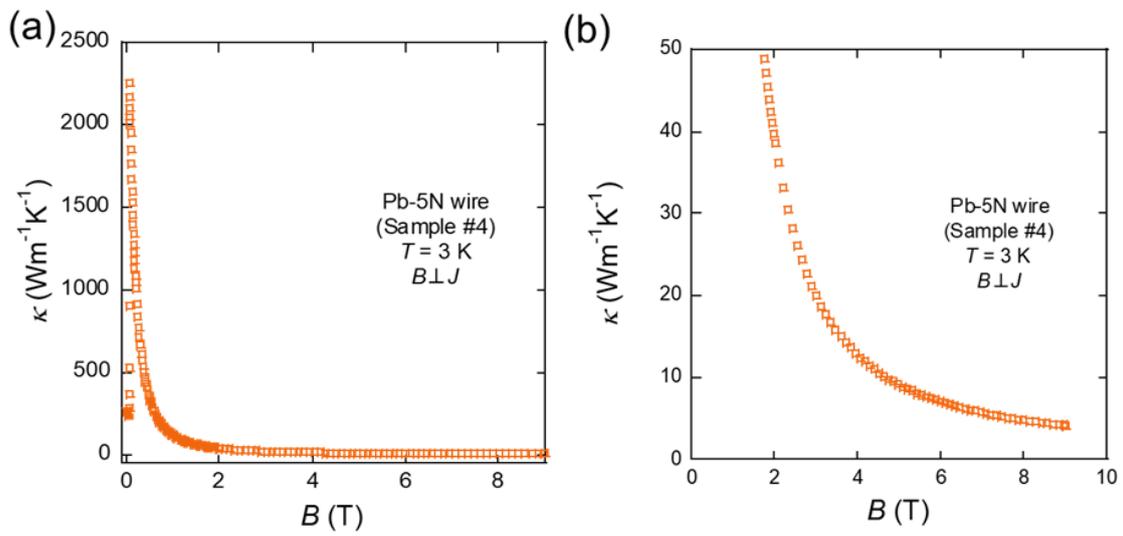

Fig. S3. *κ-B* data for Pb-5N wire (Sample #4) measured at *T* = 3 K.

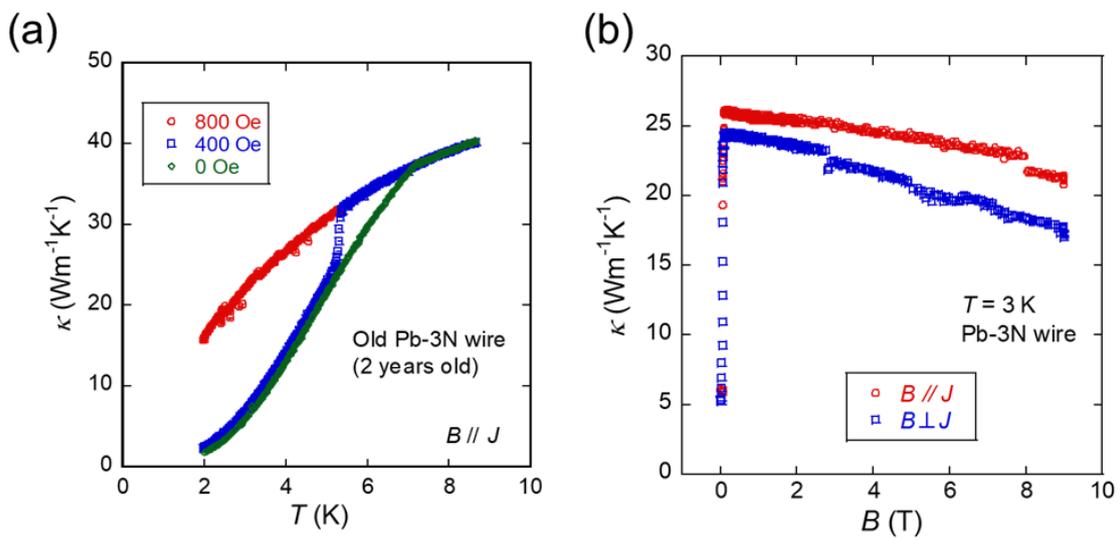

Fig. S4. (a) *κ-T* and (b) *κ-B* for Pb-3N wire.

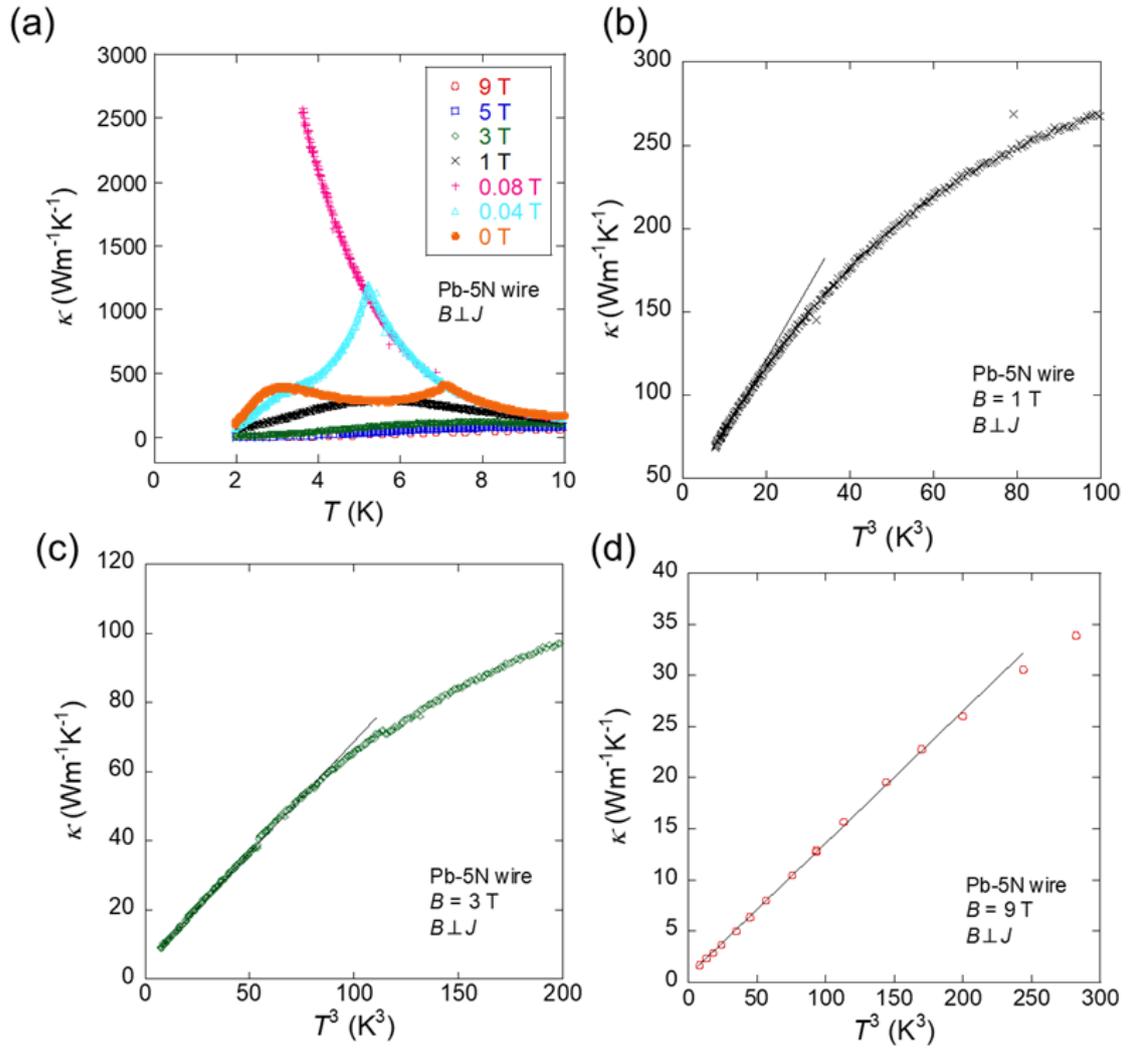

Fig. S5. (a) $\kappa$-$T$ for Pb-5N wire measured under various fields. (b-d) $\kappa$-$T^3$ plots at $B$ = 1, 3, and 9 T.

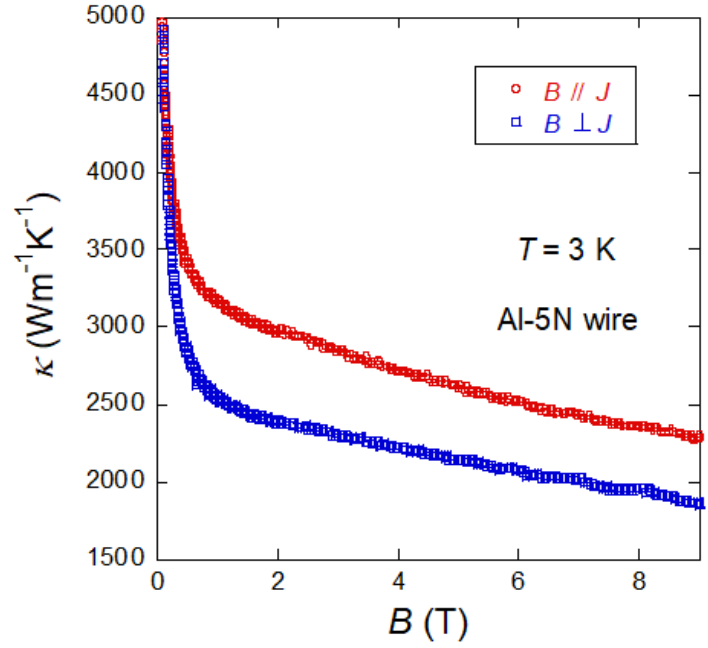

Fig. S6. *κ-B* data for Al-5N wire at *T* = 3 K.

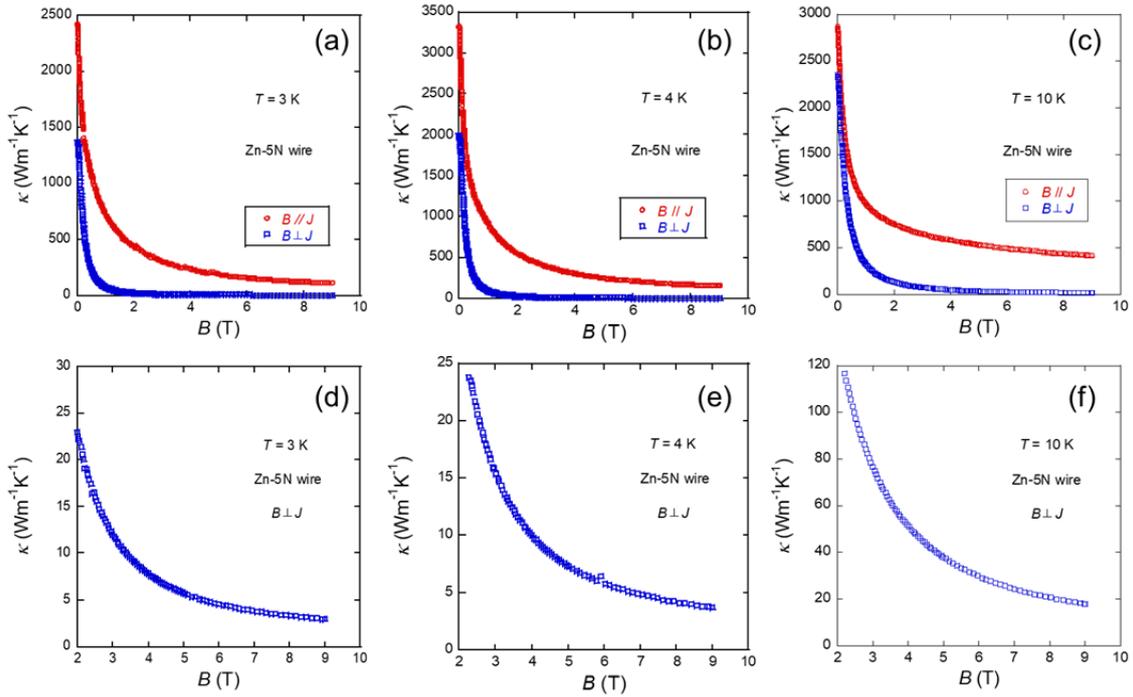

Fig. S7. (a–c) *κ-B* data for Zn-5N wire at *T* = 3, 4, and 10 K; zoomed *κ-B* are displayed in (d–f).